# Enhanced Management of Personal Astronomical Data with FITSManager

Chenzhou Cui[1*], Dongwei Fan[1], Yongheng Zhao[1], Ajit Kembhavi[2], Boliang He[1], Zihuang Cao[1], Jian Li[1], Deoyani Nandrekar[2,3]

**Affiliations**

1. National Astronomical Observatories, Chinese Academy of Sciences. 20A Datun Road, Chaoyang District, Beijing 10012, China. ccz@bao.ac.cn
2. Inter University Centre for Astronomy and Astrophysics. Post Bag 4, Ganeshkhind, Pune 411007, India. akk@iucaa.ernet.in
3. Department of Physics and Astronomy, The Johns Hopkins University, 3701 San Martin Drive, Baltimore, MD 21218 US. deoyani@pha.jhu.edu

## Abstract

Although the roles of data centers and computing centers are becoming more and more important, and on-line research is becoming the mainstream for astronomy, individual research based on locally hosted data is still very common. With the increase of personal storage capacity, it is easy to find hundreds to thousands of FITS files in the personal computer of an astrophysicist. Because Flexible Image Transport System (FITS) is a professional data format initiated by astronomers and used mainly in the small community, data management toolkits for FITS files are very few. Astronomers need a powerful tool to help them manage their local astronomical data. Although Virtual Observatory (VO) is a network oriented astronomical research environment, its applications and related technologies provide useful solutions to enhance the management and utilization of astronomical data hosted in an astronomer's personal computer. FITSManager is such a tool to provide astronomers an efficient management and utilization of their local data, bringing VO to astronomers in a seamless and transparent way. FITSManager provides fruitful functions for FITS file management, like thumbnail, preview, type dependent icons, header keyword indexing and search, collaborated working with other tools and online services, and so on. The development of the FITSManager is an effort to fill the gap between management and analysis of astronomical data.

**Keywords**: methods: data management; techniques: virtual observatory; astronomical data bases

## 1. Introduction

Astronomy depends upon observations, which are recorded with naked eyes, films or modern digital detectors. Scientific data, including observational data and other kinds of data, for example simulation results, are basic resources for astronomy research. During the last few decades, with the development of detectors and information technologies, astronomers' data harvesting abilities have increased sharply. Astronomy is facing a data deluge, and traditional data processing and analysis are facing big challenges (Szalay and Gray, 2001; Bell et al, 2009; Hey et al, 2009). As one of the solutions for these challenges, Virtual Observatory (VO) was initiated around 2000 and later an international working environment (IVOA, the International Virtual Observatory Alliance)[1] was established in 2002.

[*] National Astronomical Observatories, Chinese Academy of Sciences. 20A Datun Road, Chaoyang District, Beijing 10012, China. Tel: +86-10-64872500. FAX: +86-10-64888708. Email: ccz@bao.ac.cn

A Virtual Observatory is a data-intensive online astronomical research and education environment, taking advantages of advanced information technologies to achieve seamless, global access to astronomical information (Cui and Zhao, 2008). The power of the World Wide Web is its transparency. It is as if all the documents in the world are inside your PC. The idea of the Virtual Observatory is to achieve the same transparency for astronomical data and other related information (Quinn et al., 2004).

Scientific data available for the whole community are increasing sharply. At the same time, taking advantages of fast network and cheap storage devices, personal collections are also increasing rapidly. For an astronomer working on observational astronomy, it is very common to find hundreds or thousands of FITS files in his personal laptop. These astronomers often feel it is difficult or very hard to look for a specific file from large number of directories and files, and what is worse is that they can hardly find any tool to help them.

During the last decades, several popular data processing and analyzing packages, i.e. IRAF[2], ESO-MIDAS[3], SAOImage DS9[4], Aladin[5], fv[6], et al., have been developed to meet astronomers' requirements, but none of them provides powerful data management functions. FITS is the most common data format for astronomers (Wells et al, 1981; Hanisch et al, 2001), but it is used only in astronomy and a very few close related fields. Its user base is a very small community. For general image formats, i.e. JPG, GIF, TIF, etc, there are quite a few image management applications, freeware or commercial ones, for example ACDSee[7], Google Picasa[8], and so on. Unfortunately, none of these supports files in FITS format very well.

In professional data centers, database management systems (DBMS) and data grid technologies are used to manage large number of data files and their metadata. Finding and using such DBMS and professional data management technologies are difficult matters for most of astronomers. Thus, there is a gap between data analysis and data management for personal users. FITSManager[9] (FITS Manager) is just the tool, aiming to provide powerful and easy-of-use management functions for individual astronomers, and at the same time leveraging these functions with data analysis tools and services, including not only VO-ready ones but also legacy applications. Compared to professional data analysis packages, i.e. IRAF, MIDAS and IDL Astronomy User's Library[10], the role of the FITSManager is very similar to the role of Google Picasa to Adobe

[1] http://www.ivoa.net
[2] http://iraf.noao.edu/
[3] http://www.eso.org/sci/software/esomidas/
[4] http://hea-www.harvard.edu/RD/ds9/
[5] http://aladin.u-strasbg.fr/
[6] http://heasarc.nasa.gov/ftools/fv/
[7] http://www.acdsee.com/
[8] http://picasa.google.com/
[9] http://fm.china-vo.org/
[10] http://idlastro.gsfc.nasa.gov/

Photoshop[11].

The mission of the FITSManager is to provide individual astronomers an efficient management and utilization of large collections of local stored FITS files, bringing VO to them in a seamless and transparent way[12]. In the following sections of the paper, we first describe functions and key features of the FITSManager, and then introduce the architecture and technical highlights of the software. In the last part, we discuss future trends of astronomy data management and utilization, and summarize the development and future plan for the application.

## 2. Key Features and Functions

Professional data analysis packages and toolkits provide powerful data processing, analyzing, and visualizing functions for FITS files, while providing at best very limited management functions. Users have to manage and locate their FITS files using directories and filenames. When the number of FITS files is small, these ways are reasonable and acceptable. When an astronomer has to manage thousands to tens of thousands of FITS files in his computer, it will be hard and time consuming for him to find a specific file.

FITSManager focuses on management of small and moderately large number of FITS files hosted in astronomer's personal computer, providing index, search, preview, and external linked functions. On a recently produced laptop with 2GB or more memory, it is feasible for FITSManager to take care of thousands of FITS files. When the collection reaches to tens of thousands of files or more, data center level solutions are recommended. The main interface is shown in Figure 1. Key features and functions of the current release include:

- File search based on FITS header keywords and other metadata. Two ways are provided in the current version, general search and advanced search. For general search, a Google-like form is provided as shown in Figure 2. FITS header content and filenames in the current file list will be used to match the search phrase. Advanced search is a database based way. Before using this function, header keyword information of FITS files under a given directory or a whole file system has to be imported into a database system. Graphics User Interface (GUI) wizards are provided for both FITS header archiving and FITS file search. The file search interface and back-end programs will be improved in the future versions with guidance of user's feedback.
- Grouping, annotating and rating. The FITS files hosted in the user's computer can be grouped into different categories by their content, type or other characters. Annotation and 5-star rating can be added for selected FITS files, and these are searchable later. Being similar to functions provided by many other image management applications, all of these functions are useful for file management.
- Thumbnail and preview. FITSManager explorer provides several views for files and directories, including thumbnails, titles, icons, list and detail. In titles and icons modes, the

[11] http://www.adobe.com/products/photoshop/
[12] Virtual Astronomical Observatory Project Execution Plan. November 2010, version 1.1. p21. http://www.usvao.org/documents/Virtual_Astronomical_Observatory_Project_Execution_Plan_v1.1.pdf

FITSManager provides different icons to distinguish different FITS file types, i.e. image, random groups, multiple HDU (Header and Data Unit) FITS. In thumbnails mode, the software creates and displays thumbnails for several typical FITS format files. In the current release, table, 2D image and 1D spectrum types are supported. In future releases, thumbnail function for more types of FITS files will be provided. This function is very helpful for rapid inspection and locating of FITS files containing image and spectra data.

- Interoperability with VO tools. IVOA VOTable[13] data format and SAMP[14] messaging are supported by the FITSManager. IVOA members have recognized that building a monolithic tool that attempts to fulfill all the requirements of all users is impractical, and it is a better use of the limited resources to enable individual tools to work together efficiently. SAMP is an IVOA messaging protocol that enables astronomy software tools to interoperate and communicate. As mentioned above, FITSmanager focuses itself on file management, but leaves other functions, i.e. processing, visualization, analyze, statistics et.al, to other existing and new developed tools. Selected files in the FITSmanager can be sent to SAMP-ready applications, for example Aladin, Topcat[15], SAOImage DS9, Worldwide Telescope[16] (WWT), and so forth.
- Easy-To-use graphic user interface. FITSManager is an end user application, so a friendly user interface is very important. FITSManager explorer looks like MS Windows File Explore. An astronomer can browse his directory and file tree using the same manner as using file manager provided by his operating system (OS). When a FITS file is selected, some specific functions will be available. For files in other formats, the navigation process is similar to using an OS file manager.

## 3. Architecture and Technical Highlights

During the design and development of the FITSManager, advanced development modes were adopted and latest VO functions were considered.

### 3.1 Plug-in based architecture

Plug-in development mode is a very popular way for software development nowadays. Many well-known packages, for example Firefox[17], Eclipse[18] and many content management systems (CMS), are developed based on plug-ins.

Eclipse with Equinox[19] is used as the developing environment. Open Service Gateway Initiative (OSGi) technology is the dynamic module system for Java, which enables the modular assembly of software built with Java technology. From a code point of view, Equinox is an implementation of the OSGi[20] framework specification, a set of bundles that implement various optional OSGi

13 http://ivoa.net/Documents/VOTable/
14 http://ivoa.net/Documents/latest/SAMP.html
15 http://www.star.bris.ac.uk/~mbt/topcat/
16 http://www.worldwidetelescope.org/
17 http://www.mozilla.com/en-US/firefox/
18 http://www.eclipse.org/
19 http://www.eclipse.org/equinox/
20 http://www.osgi.org

services and other infrastructure for running OSGi technology based systems. On the other hand, Eclipse provides powerful plug-in development environment for the OSGi framework, which makes plug-in coding much easier.

In the FITSManager, all of the specific functions are realized as dynamic modules, i.e. plug-ins. A core plug-in provides common interfaces for other application plug-ins, and application plug-ins provide the dedicated functions mentioned in the above section.

In the OGSi framework, plug-in's discovery, load and execute are all completed by the framework. Comparing to the framework, all the attached plug-ins have equal roles. However, for the FITSManager package, there is a plug-in to provide basic functions. We call this plug-in as core plug-in, and other plug-ins as application plug-ins. Although the FITSManager application plug-ins are loaded by the OSGi framework physically, their functions depend on the core plug-in. Relationships among core plug-in, application plug-ins and the OGSi framework are shown in Figure 3.

When basic control functions of application plug-ins, for example, start/stop, are integrated into the core plug-in, the backend OSGi framework can be ignored in logical. Architecture of the FITSManager will be simplified as in Figure 4, consisting of a core and a bundle of application plug-ins.

FITSManager core plug-in is the kernel for the application. It controls start and stop of application plug-ins, and provides extension interfaces to them. The main functions of the core plug-in include software configuration, I/O interface management, database connection, different views of file system, etc.

Several basic application plug-ins have been provided:

- Explorer, a file manager similar to those provided by many OS distributions, for instance MS Windows and various Linux systems (Figure 1).
- Category, enable user to create custom groups for FITS files, to annotate and rate these files (Figure 5).
- Favorites, an index view of selected file system directories (Figure 6). Header information of FITS files under these directories will be imported into databases and indexed. And then fast FITS file search on FITS header keywords is enabled.
- IconCreater, return different sets of icons for current file system content to the FITSManager core plug-in according its request (Figure 7). Five types of views, thumbnails, icons, tiles, list, detail, are supported. A worthwhile point we should mention here is that the IconCreater plug-in will read and analyze header information of FITS files and create different icons for them according to their FITS types, i.e. image, multiple HDU, etc. Additionally, the plug-in also act at a FITS viewer (Figure 8).

  When the core plug-in requests a thumbnail view, the IconCreater loads the first data HDU of a FITS file into memory and return a thumbnail for the file. If values of NAXIS1 and NAXIS2 keywords in the FITS header of a file are not in the same magnitude level, for example, several vs. thousands, the IconCreater plug-in will assume the content of the file is a spectrum, and plot the first line as its thumbnail. If the FITS file is an image, a thumbnail will return after essential processes to the first data HDU, for example cutout, sharpen, and contrast adjustment. For Table FITS files, a blank table frame with defined rows and columns by NAXIS1 and NAXIS2 keywords will be returned as thumbnails.

Creating thumbnails on the fly for FITS files is a new function brought forward by the application. When an astronomer is looking for FITS files with significant characters among a number of files, this function will be very useful.

- SAMP, supports message exchange between FITSManager and other SAMP-ready tools, for example Aladin, Topcat, VOPlot, SAOImage DS9, WWT, etc. (Figure 9), which links them into an astronomical data management and utilization environment.

3.2 Other technical highlights

FITSManager is developed in Java, so operating system independence is a natural character of it. Although database systems are very important and popular for developers, they are not so familiar to astronomers, who are the end users of the package. So in the FITSManager, an embedded Java database management system, Apache Derby[21], is adopted. For embedded database component, installation and configuration are not required. When astronomers are using flexible index and search functions, they would not notice that a database system is running at the back end. If a user needs to use a standalone database system, FITSManager also provides interface to connect with it.

3.3 Performance test

Performance of two key features of the application was tested using data of one spectroscopic plate from SDSS DR7[22]. The test was conducted on a Lenovo ThinkPad T410i laptop with Intel (R) Core (TM) i5 CPU (M430 @ 2.27GHz), 4GB RAM, Microsoft Windows 7 32bit professional OS, Western Digital WD3200BEVT-08A23T1 ATA hard disk.

In the plate 1529-52930 of tile 9435, there are 640 1d spectrum FITS files at "spSpec" subdirectory and 640 image FITS files at "spAtlas" subdirectory. In "thumbnail" view mode, it takes less than one second to create thumbnails for the 640 spectrum or image FITS files in one directory.

It takes a longer time to create index for FITS files in a given directory. For the instance of the SDSS spectroscopic plate 1529-52930, it takes about 10 minutes to import FITS header content of 640 spectra files into FITSManager embedded database, which means it takes about 1s per file. When the FITS header is simpler, the process will be faster. Fortunately, this operation is not executed very often. Once an index is created, it can be used for a long time if there is no change in the directory.

---

21 http://db.apache.org/derby/
22 http://www.sdss.org/dr7/

The above results are obtained when the laptop is mainly used by the FITSManager. If there are other applications running at the same time, time needed for thumbnail creation and indexing will be longer.

## 4. **Summary**

With the arrival of Cloud computing[23], it seems that everything will appear as a service (XaaS), borders between local storage and remote storage, between local applications and on-line services become more and more blur, which is changing our way of data management and utilization. More and more research activities will take advantages of on-line services. On the other hand, personal computers and laptops will keep their role for quite a long time as the dominant working platform for a scientist, and offline research should not be ignored. FITSManager is such a tool that brings VO functions and online services to today's astronomers in a seamless and transparent way, no requirement for its user to take care of the physical locations of these functions and services.

In the future versions of the software, it will be enhanced in the following manner.

- Deeper involvement with VO and on-line services. Ground-breaking applications and on-line services appear from time to time, for example "astrometry.net" (Lang, et. al., 2010). Taking advantages of the OSGi module developing environment, plug-ins can be developed for these smart services and integrated into the FITSManager seamlessly. Plug-ins for astrometry.net, VOSpace[24] and Montage[25] are planned for the next release.
- Developer and user community. The OSGi technology provides the standardized primitives that allow applications to be constructed from small, reusable and collaborative components. These components can be composed into an application and deployed. Furthermore, dynamic module development mode enables a loose collaboration way, i.e. developer community. A FITSManager developer and user community is planned. Third party developers and end users can develop their own plug-ins and contribute to the whole community, which will enrich functions of the application rapidly.
- Enhanced thumbnail and preview functions. On one hand, more types of FITS files will be supported, for example multi-fiber spectrum, data cube, etc. On the other hand, optimized thumbnails and previews are tried to provided, which need consideration of many aspects of the data files and intelligent analysis for the content.
- Easy-of-use user interface. During the past few decades, many professional astronomers have been used to command line interfaces (CLI) coming with astronomical data process platforms, i.e. IRAF, ESO-MIDAS, and Unix like operating systems. Nowadays, network is stepping into service oriented and cloud computing era. Advanced graphic user interfaces, for example, the Multi-Touch screen from Apple iPad[26], have acquired dominant positions as user interfaces, and have been accepted widely by the younger generation. MSR Worldwide Telescope is a successful example on user interface design in astronomical software. For FITSManager, one of a number of astronomical tools, a user friendly and easy-of-use GUI is very important. In the future versions, menu layout and window theme will be redesigned. The file search dialog will be greatly improved, providing a much more friendly appearance.

[23] http://en.wikipedia.org/wiki/Cloud_computing
[24] http://www.ivoa.net/Documents/VOSpace/
[25] http://montage.ipac.caltech.edu/
[26] http://www.apple.com/ipad/design/

- Comprehensive documentation and help system. Documentation and help system are very important for users, especially for a new application. FITSManager user manual is under preparation, and a searchable "Help" button will be implemented in the tool.

**Acknowledgement**

FITS Manager is a collaborated project between Chinese Virtual Observatory (China-VO) and Virtual Observatory India (VO-India). This paper is funded by National Natural Science Foundation of China (10820002, 60920010, 90912005), Ministry of Science and Technology of China (BSDN2009-07), Beijing Municipal Science and technology Commission (2007A085). VO-India is a collaboration between the Inter-University Centre for Astronomy and Astrophysics and Persistent Systems Ltd., and is partially funded by the Ministry of Communications and Information technology of the Government of India. The first author thanks Robert Hanisch (STScI), Alex Szalay (JHU) and Mark Allen (CDS) for their helpful discussion.

**Figures**

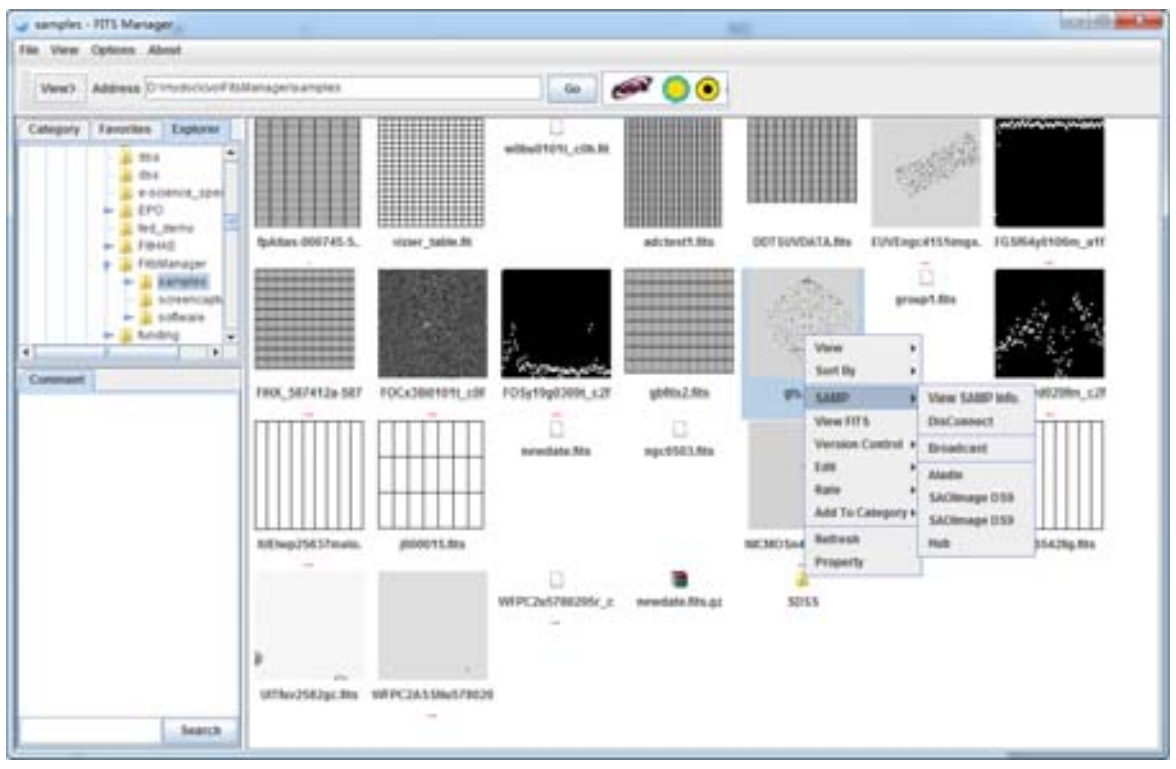

Fig.1 Main interface of FITSManager

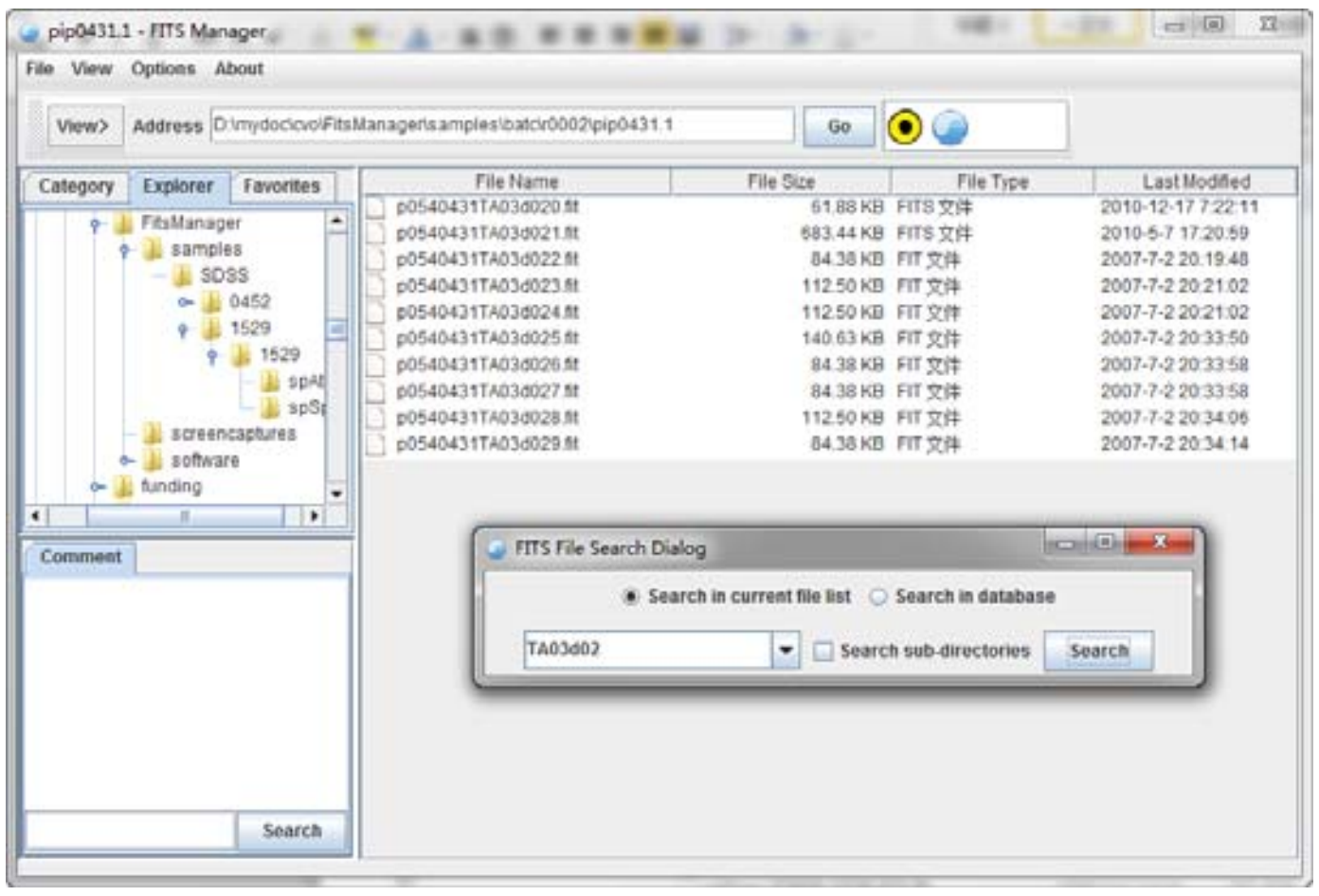

Fig. 2 File Search Dialog

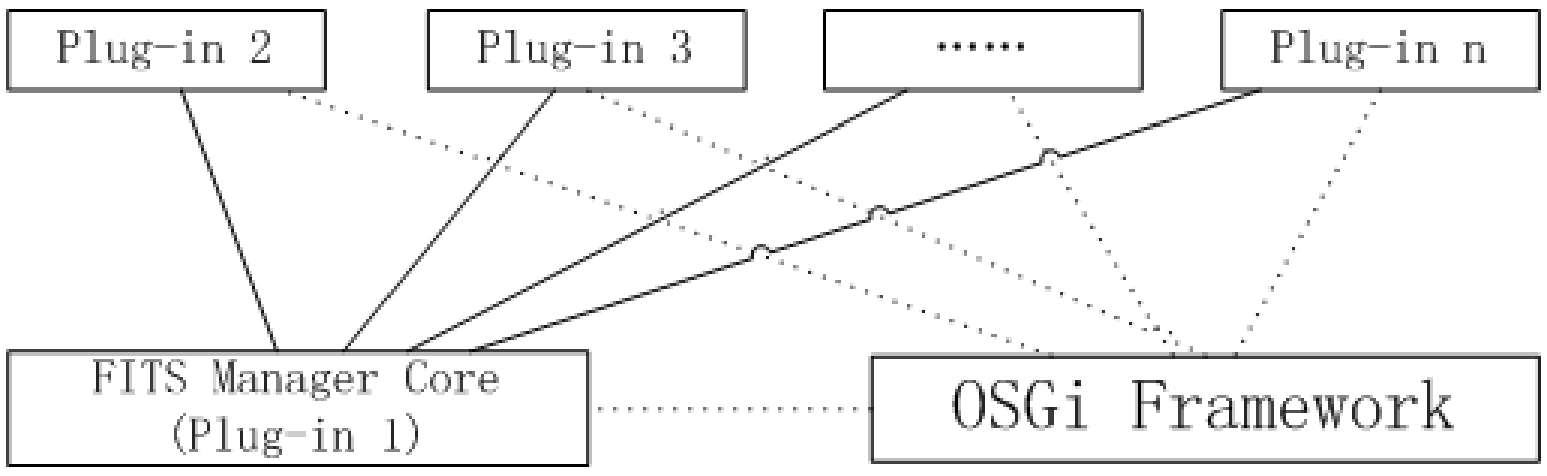

Fig.3 Relationships among core plug-in, application plug-ins and OSGi framework

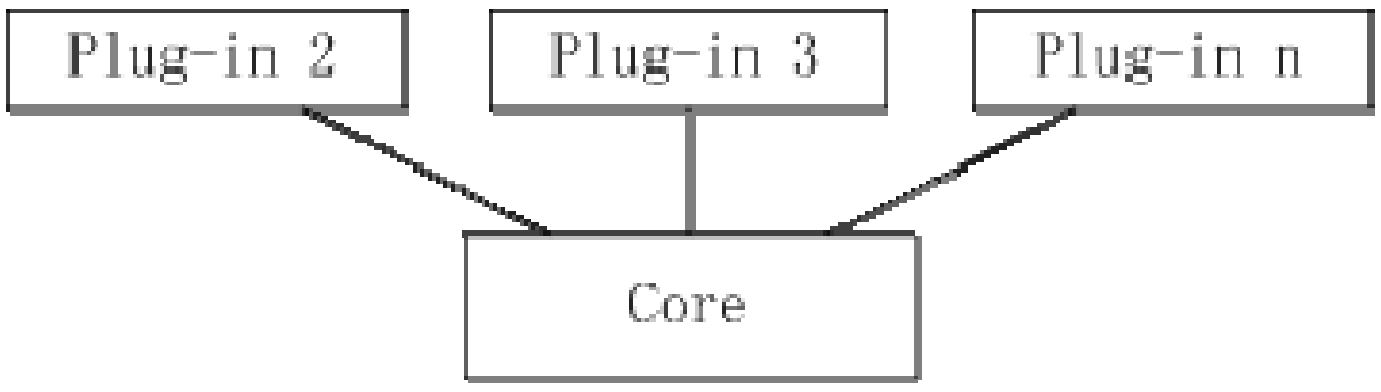

Fig.4 Logical architecture of FITSManager

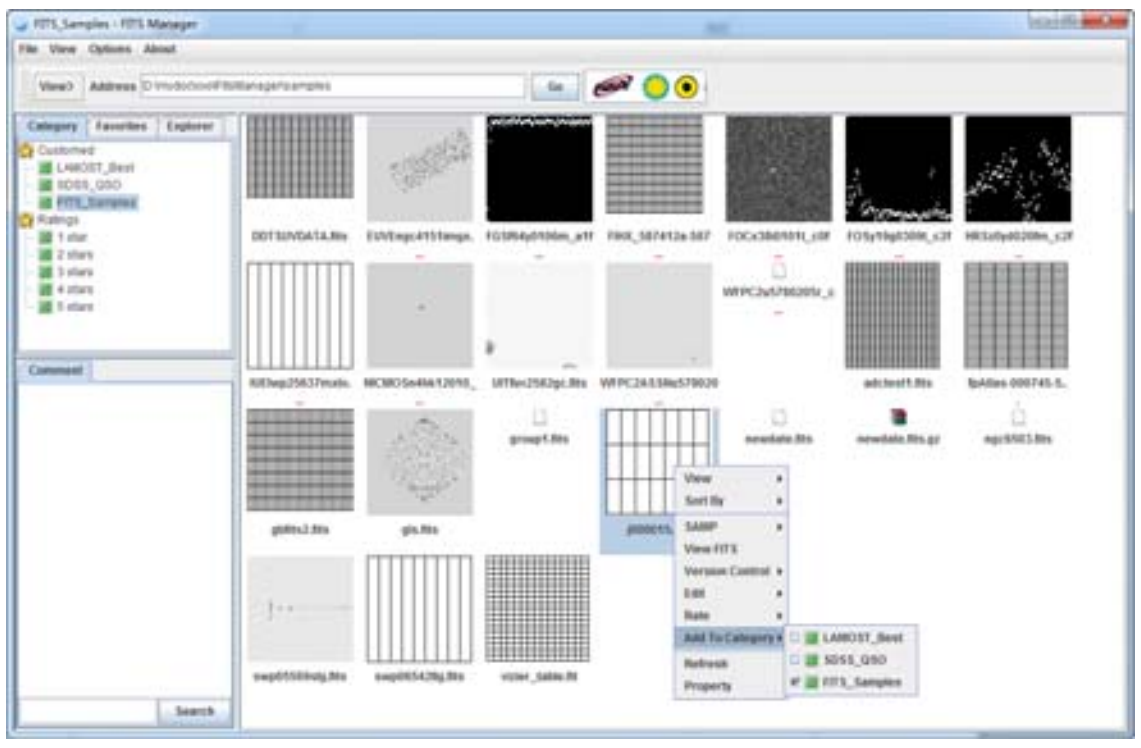

Fig.5 FITS file grouping and rating window

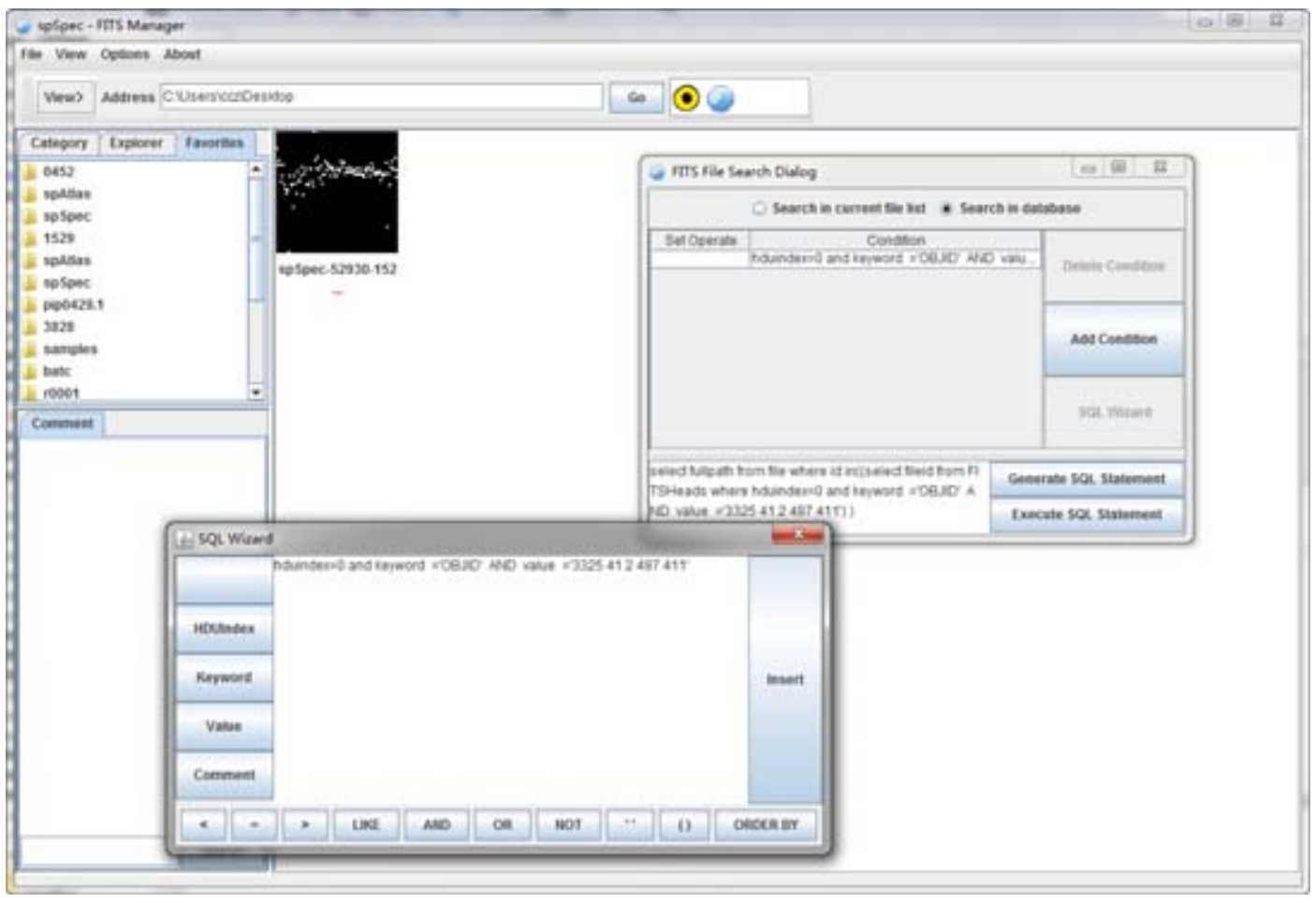

Fig.6 Index and search based on FITS header keywords

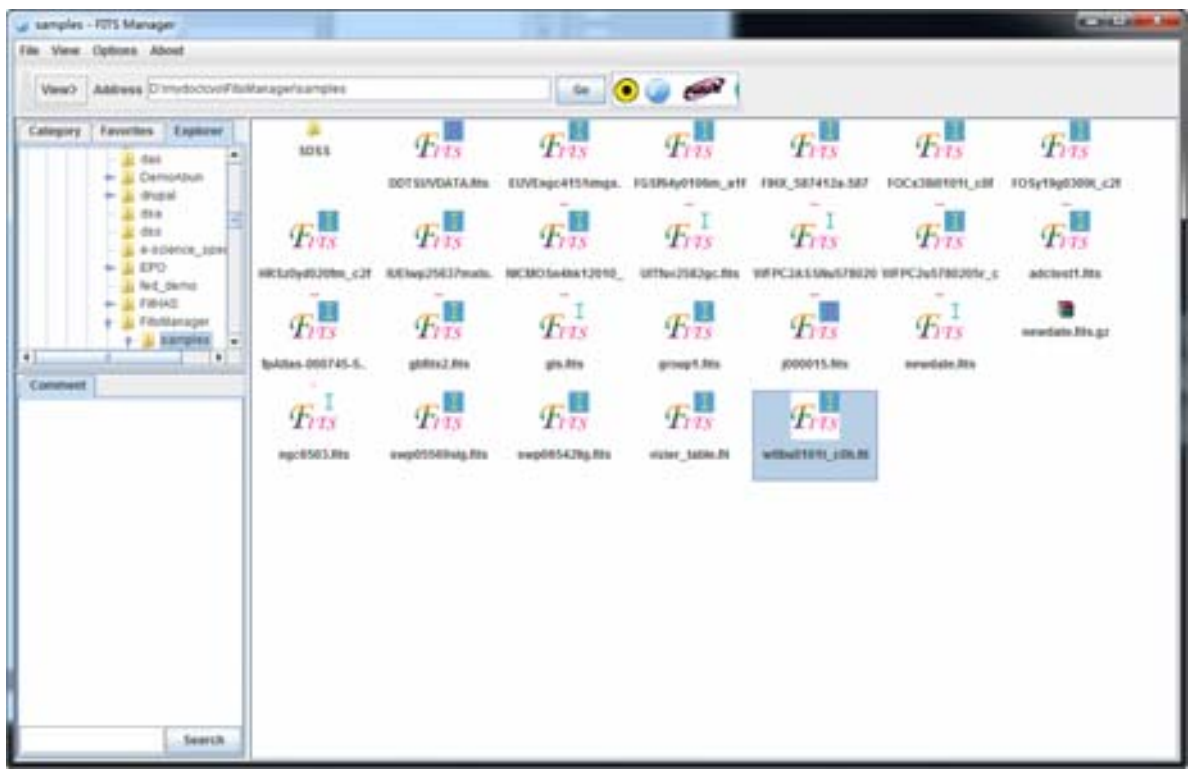

(a) Icons view

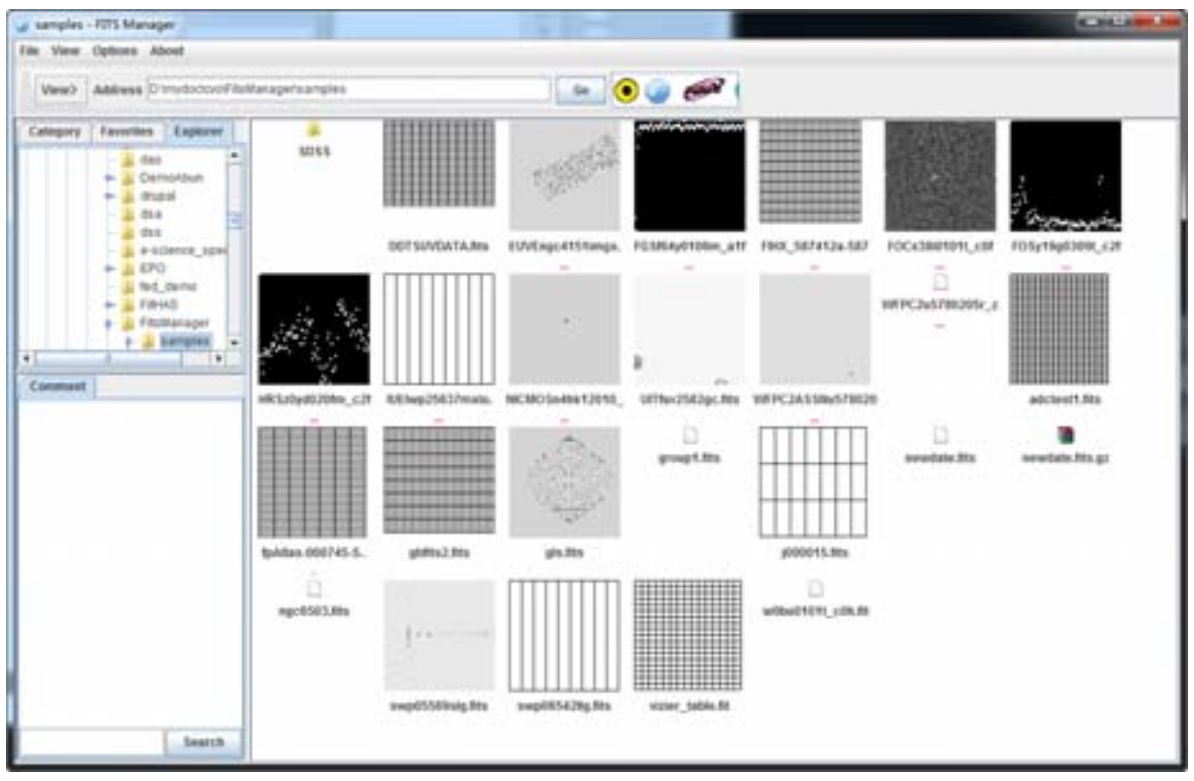

(b) Thumbnails view

Fig.7 Different views of data files, (a) shows “Icons” view and (b) shows “Thumbnails” view.

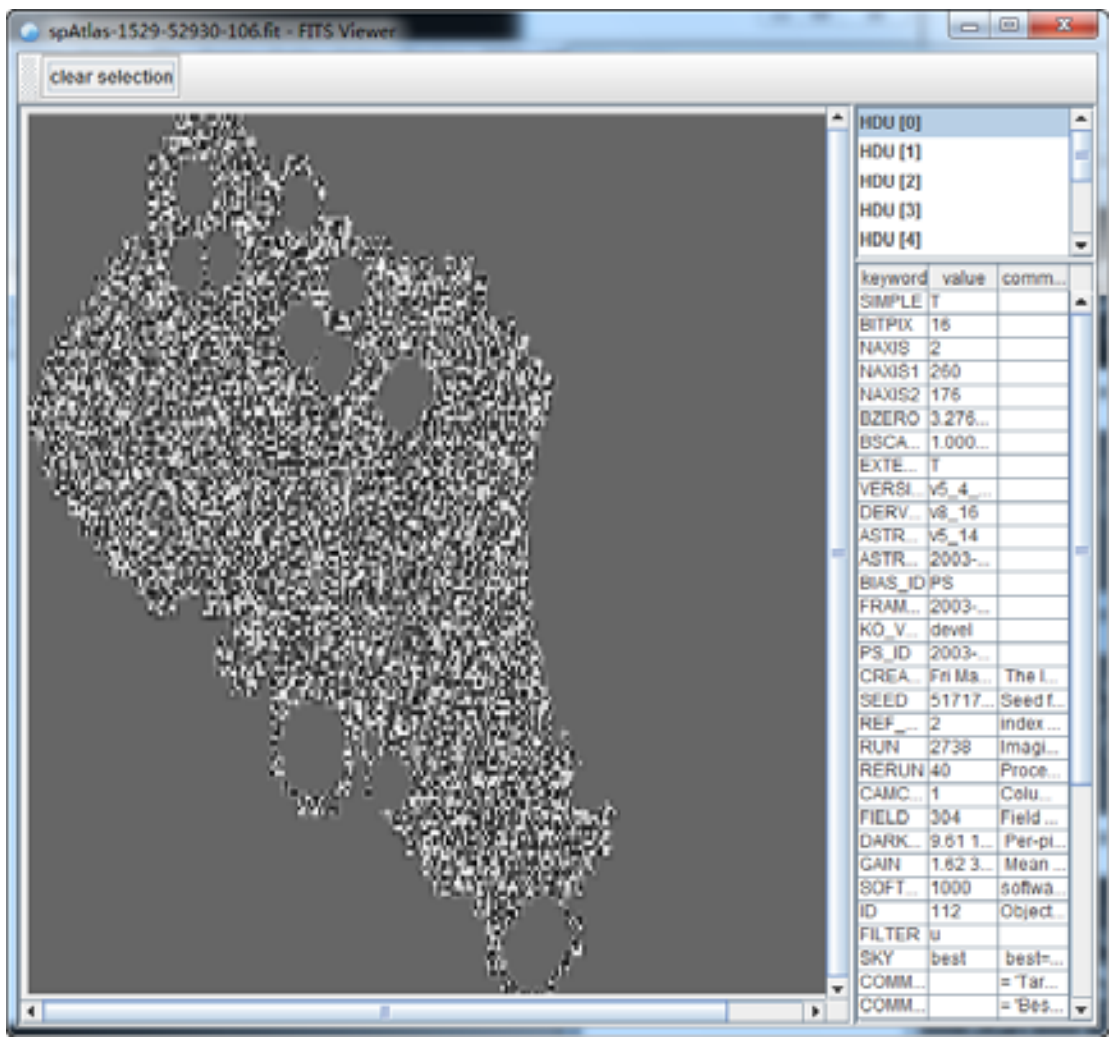

(a) Display content of a FITS file

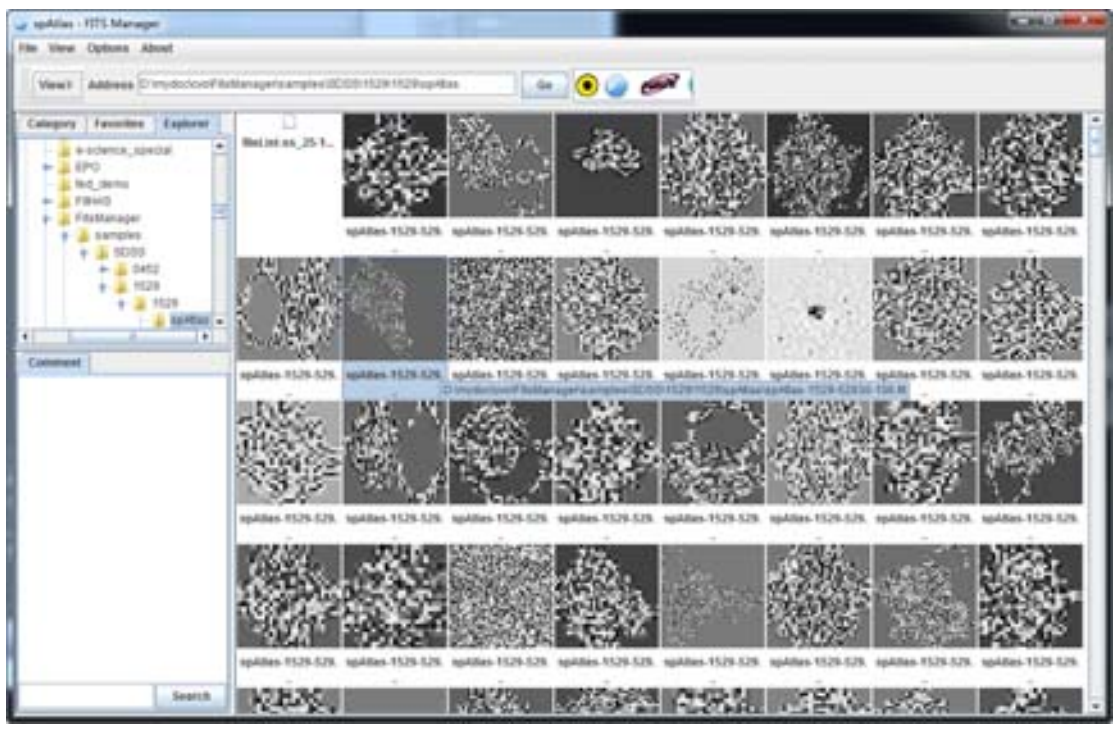

(b) Thumbnails view

Fig.8 View FITS file content and its thumbnail. Figure (a) shows the content of a selected FITS file in Figure (b).

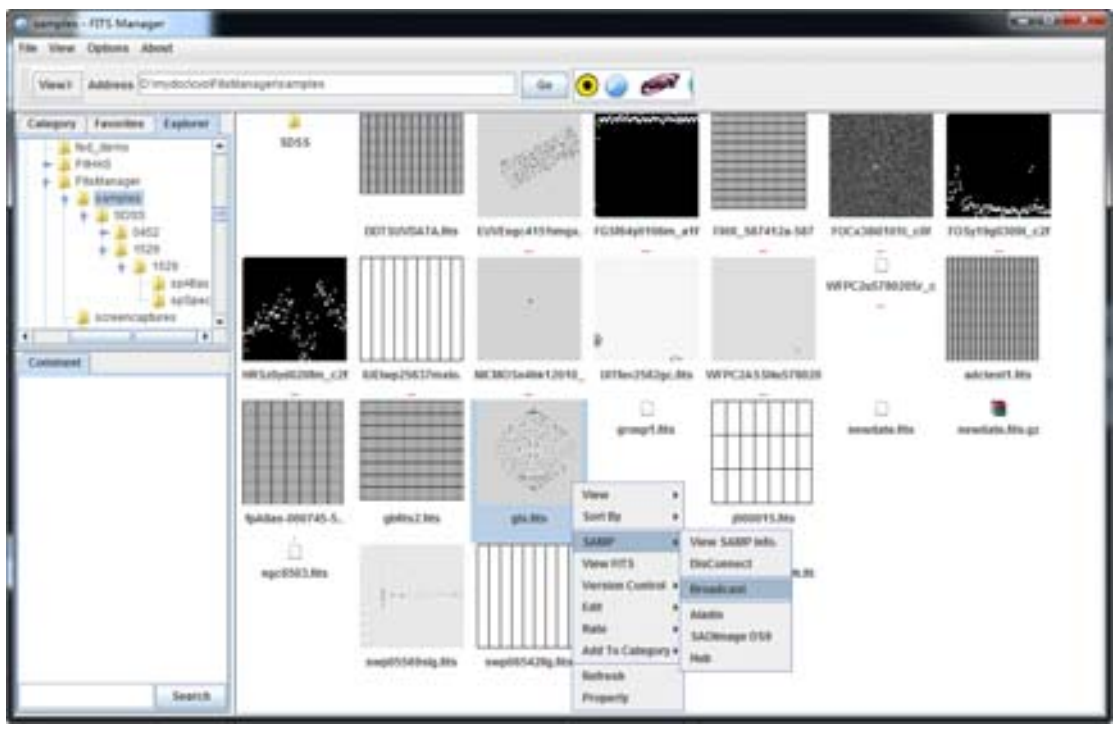

(a) Select a file and send it to other applications through SAMP

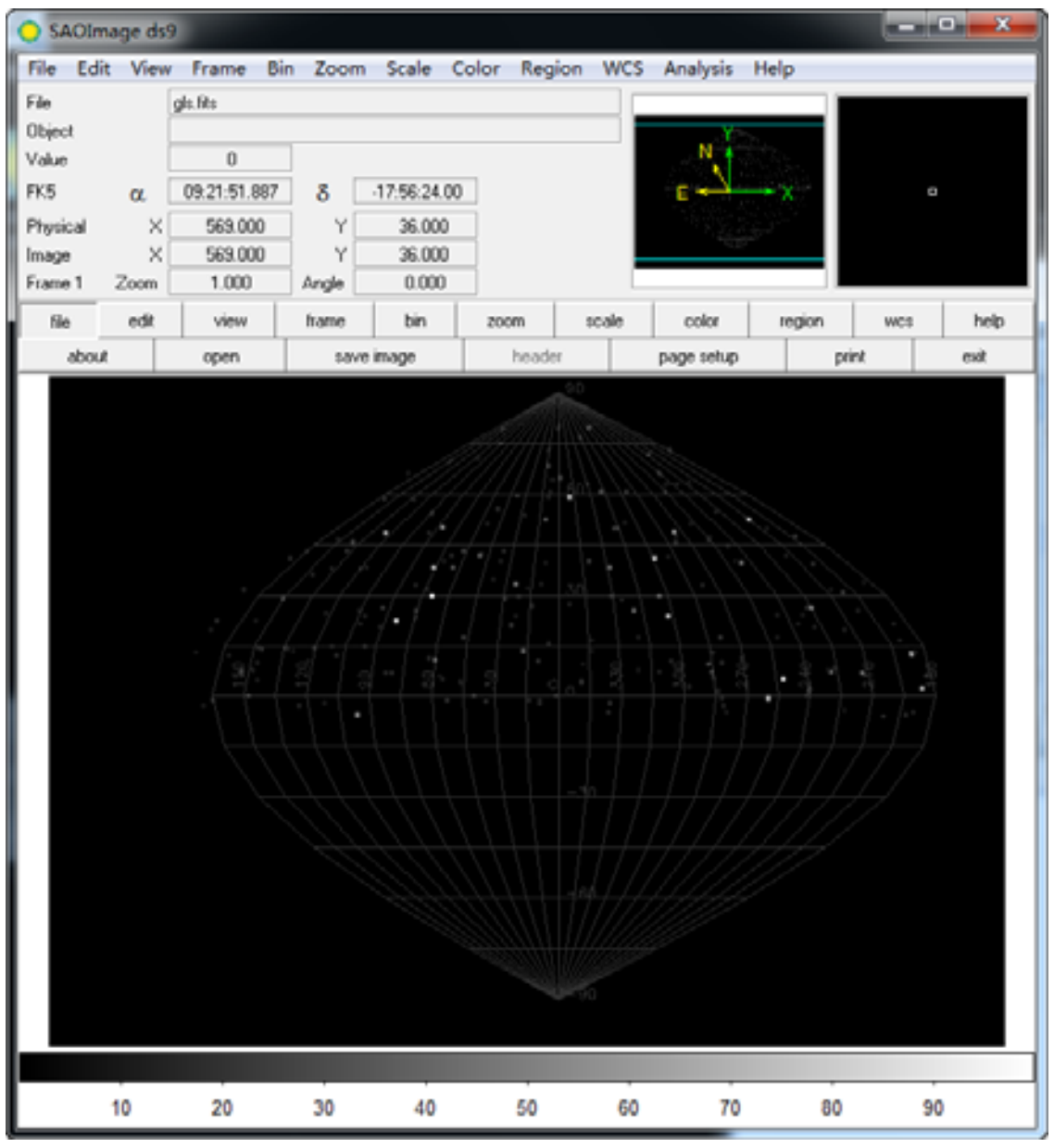

(b) File received by SAOImage DS9

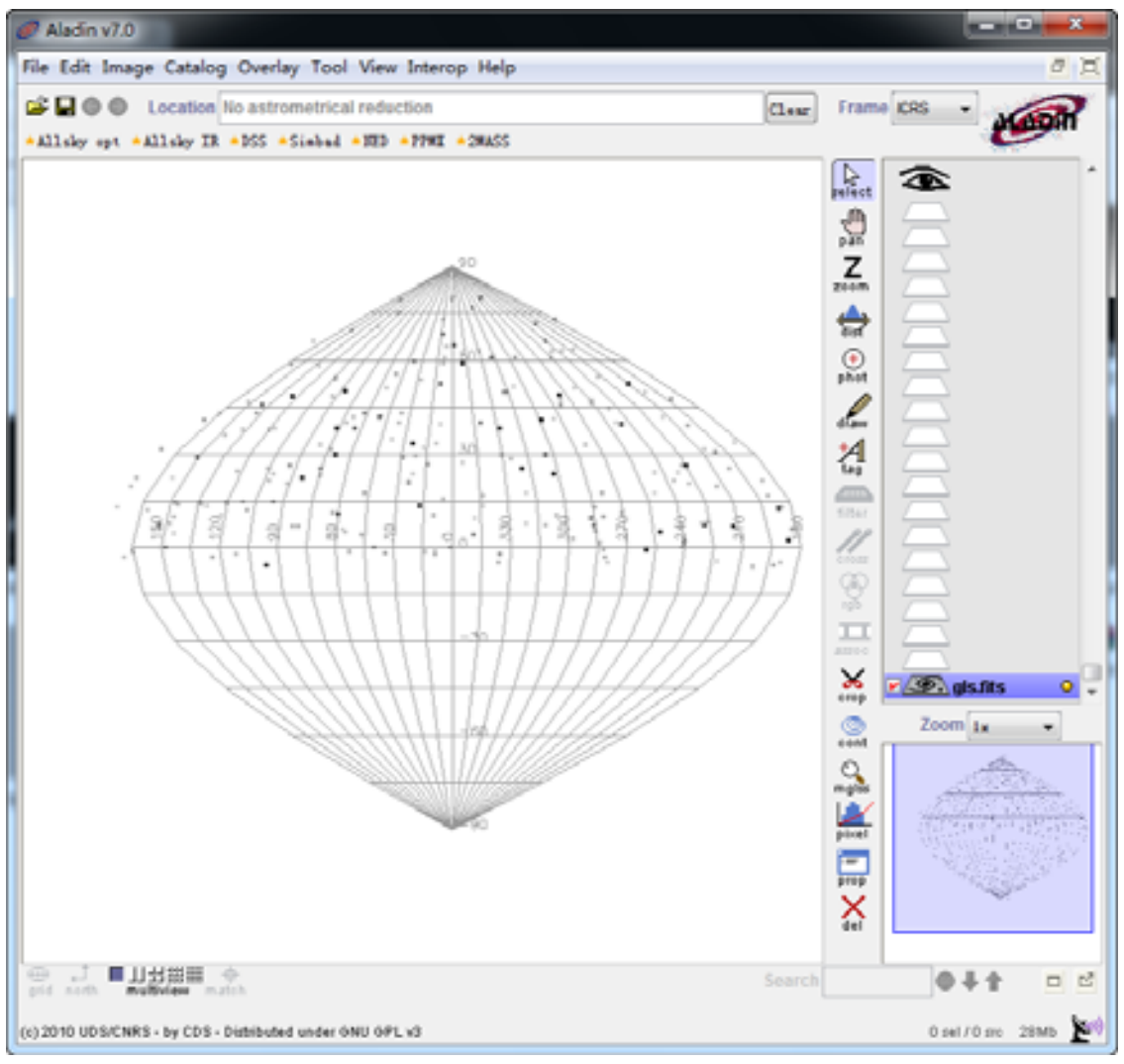

(c) File received by Aladin

Fig.9 Data exchange between FITSManager and SAOImage DS9 and Aladin